\documentclass[prd,twocolumn,superscriptaddress,nofootinbib,floatfix]{revtex4-2}
\usepackage{amsmath,amssymb,graphicx,booktabs,xcolor}
\usepackage[colorlinks,citecolor=blue!50!black,linkcolor=blue!50!black,urlcolor=blue!50!black]{hyperref}

\newcommand{\bmin}{\beta_{\rm min}}
\newcommand{\tOm}{\tilde\Omega}

\begin{document}

\title{Coupled dark energy and apparent phantom crossing: 
a potential-independent lower bound}

\author{Domenico Sapone}
\email{domenico.sapone@uchile.cl}
\affiliation{Departamento de F\'isica, FCFM, Universidad de Chile, Blanco Encalada 2008, Santiago, Chile}

\author{Esteban Arratia}
\email{esteban.arratia@ug.uchile.cl}
\affiliation{Departamento de F\'isica, FCFM, Universidad de Chile, Blanco Encalada 2008, Santiago, Chile}

\date{\today}

\begin{abstract}

Baryon acoustic oscillation measurements from the Dark Energy Spectroscopic Instrument (DESI), combined with the cosmic microwave background (CMB) and type Ia supernovae, prefer a dark energy whose equation of state was below $-1$ in the past and is above $-1$ today.
A canonical scalar field cannot cross this line, but it can appear to do so if it is coupled to dark matter, and we ask how strong this coupling must be, for any potential.
We show that the apparent crossing requires the dark matter to lose a minimum fraction of its mass after recombination, fixed by the expansion history $H(z)$.
For a constant coupling $\beta=d\ln m/d\phi$, the mass can change only as fast as the field moves, and the speed of the field is itself set by $H(z)$, so that the coupling must be larger than a minimum value $\bmin[H(z)]$.
This bound holds for any potential, applies to the largest value of the coupling when the coupling depends on the field, and is reached exactly when the field comes to rest within the observed range.
For the DESI DR2 best fits, the dark matter must lose about 5--7\% of its mass by $z\simeq0.8$, which requires $\bmin=0.047$--$0.058$; including the uncertainties of the data, the public DESI DR2 chains give the 95\% lower limit $|\beta|>0.025$--$0.038$.
We also apply the bound to published fits of coupled quintessence: all of them respect it, and those with a constant coupling reproduce only a shallow version of the crossing.
Since the coupling cannot be smaller than $\bmin$, two known effects of the coupling have a minimum size: an extra attraction between dark matter particles of at least a few tenths of a per cent of gravity, and a lensing signal at least a few per cent weaker than the clustering of dark matter would suggest.
\end{abstract}

\maketitle

%=======================================================================
\section{Introduction}
\label{sec:intro}

The second data release of the Dark Energy Spectroscopic Instrument (DESI), combining baryon acoustic oscillations (BAO) with the cosmic microwave background (CMB) and type Ia supernovae (SN), prefers a dark energy that evolves with time~\cite{DESI:2025zgx,DESI:2025fii}.
In the $w_0w_a$ parametrisation the equation of state lies below $-1$, in the phantom regime, at $z\gtrsim0.4$, and above $-1$ today, at a significance of $\sim 3\sigma$; the low-redshift crossing $w=-1$ is also found when the parametrisation is relaxed~\cite{Artola:2025zzb}.
Such a crossing cannot be obtained from the simplest dynamical dark energy, a canonical scalar field, because its energy density and pressure satisfy $\rho_\phi+p_\phi=\dot\phi^2\geq0$, so that its equation of state never goes below $-1$~\cite{Copeland:2006wr,Sapone:2010iz,Huterer:2017buf}. More generally, a single dark-energy component that crosses $w=-1$ faces instabilities in its perturbations, unless additional degrees of freedom are introduced~\cite{Vikman:2004dc,Hu:2004kh,Kunz:2006wc}.

A coupling between dark energy and dark matter offers such an additional degree of freedom without any instability, because the crossing is only apparent.
Das, Corasaniti and Khoury~\cite{Das:2005yj} pointed out that a canonical field can appear phantom if cold dark matter (CDM) transfers energy to it: an observer who assumes that the two components are separate assigns this energy to dark energy, which then seems to grow with time, even though the field itself always has $w\geq-1$.
The idea has received new attention after the DESI results. While some works look for the origin of the signal in the dark matter sector alone~\cite{Giani:2025hhs,Braglia:2025gdo,Chen:2025wwn}, many others have studied coupled quintessence in light of the new data~\cite{Gomez-Valent:2026ept,Babu:2026tfr,Wang:2026gvg,Wang:2026vqw,Li:2026xaz,Samanta:2025oqz,Chakraborty:2025syu,Pompeu:2026zad,Antusch:2026ldp,Wang:2026wrk,deSouza:2026qyf,Khoury:2025txd,Khoury:2026svx,LaPenna:2026avs,Bedroya:2025fwh}.
These works generally find couplings $\beta=d\ln m/d\phi$ of a few per cent, that is, a change of the dark matter mass of a few per cent for a field displacement of one Planck mass, but each of them relies on a specific potential and a specific coupling function, and the fits reproduce the crossing to very different degrees.

Here we ask a question that does not depend on these choices: given an expansion history with an apparent crossing, how strong must the coupling be?
The answer follows from two simple facts.
The first is that the kinetic energy of the field cannot be negative.
The effective dark energy can then grow only if the dark matter has lost part of its mass, and the faster it grows, the larger this loss must be, so that the expansion history $H(z)$ fixes a minimum mass loss since recombination.
The second fact is that, for a constant coupling, the dark matter mass changes only when the field moves, and the speed of the field is itself set by $H(z)$.
The mass loss therefore has a speed limit proportional to the coupling: if the coupling is too small, the dark matter cannot lose enough mass in time, whatever the potential is.
This gives a minimum coupling $\bmin[H(z)]$.
If the coupling depends on the field, it may be small at recombination, but it must reach at least $\bmin$ before the crossing.
Finally, the bound is sharp: if the field comes to rest at some redshift within the observed range, the coupling is exactly $\bmin$, and for the DESI backgrounds this happens at $z\simeq0.8$, where the required loss is largest.

For DESI DR2 we find $\bmin\simeq0.05$. The requirement is larger, $0.06$--$0.08$, if the field is frozen before the data range, as in thawing models, because the field then has less time to roll.
As a consequence, two known effects of the coupling acquire a minimum size: a fifth force between dark matter particles, and a deficit of lensing relative to the clustering of dark matter.

We describe the model and the point of view of the observer in Sec.~\ref{sec:model}, and derive the bounds in Sec.~\ref{sec:bound}.
Sec.~\ref{sec:results} applies them to DESI DR2 and Sec.~\ref{sec:obs} discusses the observable consequences.
Only then, in Sec.~\ref{sec:compare}, do we compare with other works, so that our results can be read independently of any particular model.

%=======================================================================
\section{Coupled quintessence seen by an uncoupled observer}
\label{sec:model}

We use reduced Planck units, $8\pi G=1$, so that $3H^2=\rho_{\rm tot}$ and $\Omega_i=\rho_i/3H^2$.
A dot denotes a derivative with respect to cosmic time, a prime a derivative with respect to the number of e-folds $N=\ln a$, and the subscript $*$ denotes recombination.

The dark sector contains a canonical scalar field $\phi$ with potential $V(\phi)$, and CDM particles whose mass $m(\phi)$ depends on the field, while baryons and radiation are not coupled. The action is
\begin{align}
S=\int d^4x\sqrt{-g}\,\Bigl[&\frac{R}{2}-\frac12 g^{\mu\nu}\partial_\mu\phi\,\partial_\nu\phi\nonumber\\
&-V(\phi)-m(\phi)\,n\Bigr]+S_{b,r}\,,
\end{align}
where $n$ is the CDM number density. Since the number of particles is conserved, $n\propto a^{-3}$, and the CDM energy density $\rho_c=m(\phi)\,n$ changes only through the mass; here and in the following, the subscript $c$ denotes CDM (not the critical density).
We define the coupling as the logarithmic slope of the mass,
\begin{equation}
\beta(\phi)\equiv\frac{d\ln m}{d\phi}\,.
\label{eq:beta}
\end{equation}
Differentiating $\rho_c=m\,n$ with respect to time, and varying the action with respect to $\phi$ in a flat FLRW background, we obtain
\begin{align}
\dot\rho_c+3H\rho_c&=\beta\dot\phi\,\rho_c\,, \label{eq:cont_c}\\
\ddot\phi+3H\dot\phi+V_{,\phi}&=-\beta\rho_c\,, \label{eq:KG}
\end{align}
and multiplying Eq.~\eqref{eq:KG} by $\dot\phi$, with $\rho_\phi=\dot\phi^2/2+V$ and $p_\phi=\dot\phi^2/2-V$, we find
\begin{equation}
\dot\rho_\phi+3H(\rho_\phi+p_\phi)=-\beta\dot\phi\,\rho_c .
\label{eq:cont_phi}
\end{equation}
The energy lost by CDM is gained by the field, at a rate $Q=\beta\dot\phi\rho_c$. A constant $\beta$ corresponds to an exponential mass: with $\beta=0.05$, a field displacement of one Planck mass changes the mass by 5\%.
Since only the product of $\beta$ and the displacement of the field is physical, we quote $|\beta|$.

An observer who assumes uncoupled components, such as the $w_0w_a$ fits of DESI, fixes the CDM density with the CMB and then dilutes it as $a^{-3}$,
\begin{equation}
\tilde\rho_c(a)=\rho_c(a_*)\left(\frac{a_*}{a}\right)^3\,,
\end{equation}
where we set $\rho_c(a_*) = \tilde\rho_c(a_*)$. In the true model, integrating Eq.~\eqref{eq:cont_c} from recombination gives $a^3\rho_c\propto m(\phi)$, so the true density is $\rho_c=[m(\phi)/m(\phi_*)]\,\tilde\rho_c$: the two densities differ only because the mass has changed.
We measure this change with the fractional mass loss since recombination,
\begin{equation}
\Delta(N)\equiv1-\frac{m(\phi)}{m(\phi_*)}=1-\frac{\rho_c}{\tilde\rho_c}\,,\qquad\Delta(N_*)=0\,,
\label{eq:Delta}
\end{equation}
which is positive when CDM has lost mass.
The observer assigns everything else in the dark sector to an effective dark energy,
\begin{equation}
\rho_{\rm eff}\equiv\rho_\phi+\rho_c-\tilde\rho_c=\rho_\phi-\Delta\,\tilde\rho_c \,,
\label{eq:rhox}
\end{equation}
so that $3H^2=\rho_r+\rho_b+\tilde\rho_c+\rho_{\rm eff}$ describes the same expansion history as the true model.
When Eq.~\eqref{eq:rhox} is inserted into Eq.~\eqref{eq:cont_phi}, the term $\dot{\tilde\rho}_c+3H\tilde\rho_c$ vanishes and $\dot\rho_c+3H\rho_c$ cancels the interaction, so that the effective dark energy is conserved,
\begin{equation}
\dot\rho_{\rm eff}+3H(1+w_{\rm eff})\rho_{\rm eff}=0\,,\qquad w_{\rm eff}=\frac{p_\phi}{\rho_{\rm eff}}\,.
\end{equation}
This $w_{\rm eff}$ is what a $w_0w_a$ fit measures. Since $\rho_{\rm eff}<\rho_\phi$ when $\Delta>0$, the effective equation of state can lie below $-1$ while $w_\phi\geq-1$.

We normalise $\tilde\rho_c$ at recombination because an analysis that includes the CMB fixes the CDM density through the acoustic peaks. A normalisation today, as in~\cite{Das:2005yj,Rosenfeld:2007ri}, would only change $\Delta$ by a constant factor, $1-\Delta_{\rm today}(z)=[1-\Delta(z)]/[1-\Delta(0)]$.

%=======================================================================
\section{A minimum coupling for any potential}
\label{sec:bound}

\subsection{A minimum loss of dark matter mass}
\label{sec:Dmin}

Since the pressure of $\rho_{\rm eff}$ is $p_\phi$, Eq.~\eqref{eq:rhox} gives $(1+w_{\rm eff})\rho_{\rm eff}=(\rho_\phi+p_\phi)-\Delta\,\tilde\rho_c=\dot\phi^2-\Delta\,\tilde\rho_c$.
Dividing by $3H^2$ and using $\dot\phi=H\phi'$, we obtain the identity
\begin{equation}
\frac{\phi'^2}{3}=\Omega_\phi(1+w_\phi)=\mathcal{B}(z)+\tOm_c(z)\,\Delta(z)\,,
\label{eq:identity}
\end{equation}
where $\tOm_c=\tilde\rho_c/3H^2$ and
\begin{equation}
\mathcal{B}\equiv(1+w_{\rm eff})\,\Omega_{\rm eff}\,.
\label{eq:Bdef}
\end{equation}
Equation~\eqref{eq:identity} is equivalent to the relation between $w_{\rm eff}$ and $w_\phi$ derived in Ref.~\cite{Das:2005yj}.
The left-hand side belongs to the true model, while the right-hand side contains, apart from $\Delta$, only quantities measured by the observer.
In fact, $\mathcal{B}$ can be written in terms of $H(z)$ alone: differentiating $3H^2=\rho_r+\rho_m+\rho_{\rm eff}$ with respect to $\ln a$, with $\rho_m=\rho_b+\tilde\rho_c\propto a^{-3}$, $\rho_r\propto a^{-4}$ and the conservation of $\rho_{\rm eff}$, we find
\begin{equation}
\mathcal{B}(z)=-\frac23\frac{d\ln H}{d\ln a}-\frac43\Omega_r(z)-\tOm_m(z)\,,
\label{eq:BH}
\end{equation}
with $\tOm_m=\tilde\omega_m a^{-3}/h^2(a)$. The first term is the measured shape of $H(z)$, and the other two require the matter and radiation densities assumed by the observer.
Since $\mathcal{B}$ depends only on $H(z)$, it can be obtained from distance data without assuming a dark-energy model, as in recent model-independent analyses of DESI DR2 BAO~\cite{Sapone:2026iwc, Martinelli:2026wjp}.
As a check, Eq.~\eqref{eq:BH} gives $\mathcal{B}=0$ for $\Lambda$CDM and $\mathcal{B}=(1+w)\,\Omega_{\rm DE}$ for $w_0w_a$; $\mathcal{B}$ is negative exactly where $w_{\rm eff}<-1$.

The left-hand side of Eq.~\eqref{eq:identity} is proportional to the kinetic energy of the field and it cannot be negative.
The right-hand side must therefore be non-negative too, which gives our first result:
\begin{equation}
\Delta(z)\;\geq\;\Delta_{\min}(z)\equiv-\frac{\mathcal{B}(z)}{\tOm_c(z)}\,.
\label{eq:Dmin}
\end{equation}
Physically, where $w_{\rm eff}<-1$ the observer sees a negative $\mathcal{B}$, which a canonical field cannot produce by itself;
the only way to compensate it is a positive $\tOm_c\,\Delta$, that is, dark matter that has lost mass.
Therefore, in this region, the dark matter must have lost at least a fraction $\Delta_{\min}>0$ of its mass since recombination, and this fraction comes directly from the observed expansion.
Where $w_{\rm eff}\geq-1$, instead, $\Delta_{\min}\leq0$: at these redshifts the expansion does not require a mass loss, although the mass lost at earlier times may still be missing.
The argument uses only energy conservation in the dark sector and $w_\phi\geq-1$, i.e.\ the middle term of Eq.~\eqref{eq:identity};
therefore it holds for any potential and any coupling function, and also for dark-energy fluids with $w\geq-1$ that exchange energy with dark matter.

It is useful to see what $\Delta_{\min}$ measures. Since $\rho_{\rm eff}$ obeys the continuity equation without sources, $(1+w_{\rm eff})\rho_{\rm eff}=-\tfrac13\,d\rho_{\rm eff}/dN$, and Eq.~\eqref{eq:Dmin} can be written as
\begin{equation}
\Delta_{\min}=\frac{1}{3\,\tilde\rho_c}\,\frac{d\rho_{\rm eff}}{dN}\,.
\label{eq:Dminslope}
\end{equation}
The minimum mass loss is therefore set by how fast the effective dark energy grows, compared with the CDM density, and not by how much it has grown.
It vanishes at the crossing $w_{\rm eff}=-1$, where $\rho_{\rm eff}$ reaches its maximum and stops growing, and it is largest at an earlier time, where this relative growth is fastest.
The strongest requirement therefore comes not from the crossing itself, but from the epoch before it.
We call the peak value of $\Delta_{\min}$ the depth of the crossing: it measures how far the effective equation of state goes below $-1$, weighted by the dark-energy fraction, and it is the largest mass loss that the crossing requires.

\subsection{A minimum coupling}
\label{sec:bmin}

Equation~\eqref{eq:Dmin} says how much mass the CDM must lose, but not how fast it can lose it.
To answer this second question we need the coupling.
For a constant coupling, Eq.~\eqref{eq:beta} admits the solution $m(\phi)=m(\phi_*)\,e^{\beta(\phi-\phi_*)}$, so that
\begin{equation}
    \Delta=1-e^{\beta(\phi-\phi_*)}\,,
\end{equation} and its derivative is
\begin{equation}
\Delta'=-\beta\,\phi'\,(1-\Delta)\,.
\end{equation}
The mass changes only because the field moves, and it changes faster when the field moves faster.
The speed of the field, in turn, is fixed by Eq.~\eqref{eq:identity}: $|\phi'|=\sqrt{3K}$, with $K\equiv\mathcal{B}+\tOm_c\,\Delta$.
Combining the two, the mass history obeys
\begin{gather}
\Delta'=\pm F(N,\Delta),\qquad\Delta(N_*)=0,\nonumber\\
F\equiv|\beta|\,(1-\Delta)\sqrt{3\bigl[\mathcal{B}(N)+\tOm_c(N)\,\Delta\bigr]}\,.
\label{eq:Dode}
\end{gather}
The sign is that of $-\beta\phi'$: it is positive when the mass decreases, and it can change only where $K=0$, that is, where the field is at rest. All the coefficients of Eq.~\eqref{eq:Dode} are background quantities.
For a given $H(z)$ and $\beta$, it therefore gives the mass history, and from it the field trajectory and the potential that reproduce exactly this $H(z)$.
Each allowed coupling describes the same expansion history with a different potential, an example of the dark degeneracy~\cite{Kunz:2006ca, Kunz:2007rk}.

However, not every coupling is allowed, because $K$ must stay non-negative wherever $H(z)$ is measured.
Since $\Delta'$ is either $+F$ or $-F$, depending only on the direction in which the field rolls, we always have $-F\leq\Delta'\leq F$: the mass loss can never grow faster than $F$, which therefore acts as a speed limit.

More precisely, let $\Delta_{\max}(N;\beta)$ be the solution that always moves at this limit, i.e. $\Delta'_{\max}=+F$, starting from $\Delta(N_*)=0$.
No other solution can overtake it.
Since both curves start from $\Delta=0$ at recombination, a solution that ended up above $\Delta_{\max}$ would first have to meet it, i.e.\ take the same value at the same time.
At that point $F$, which depends only on $N$ and $\Delta$, is the same for both curves, and since $\Delta'\leq F=\Delta'_{\max}$, the other solution cannot grow faster than $\Delta_{\max}$ and move above it.
Hence $\Delta(N)\leq\Delta_{\max}(N;\beta)$ for every potential.

Every mass history must therefore lie below $\Delta_{\max}$, while the data require it to lie above $\Delta_{\min}$.
If at some redshift $\Delta_{\max}<\Delta_{\min}$, no curve can satisfy both conditions, and no potential can produce the observed $H(z)$ with this coupling.
Since $F$ is proportional to $|\beta|$, a stronger coupling gives a higher $\Delta_{\max}$: if a coupling is allowed, every larger one is allowed too.
We therefore define the minimum coupling $\bmin[H(z)]$ as the smallest $|\beta|$ for which $\Delta_{\max}\geq\Delta_{\min}$ at all measured redshifts.
Neither $\Delta_{\min}$ nor $F$ contains the potential: $\bmin$ depends only on the expansion history after recombination and on the CDM density fixed by the CMB (see Sec.~\ref{sec:early} for the epoch before the data).

The same argument applies when $\beta$ depends on the field.
The speed limit is then at most $|\beta|_{\max}(1-\Delta)\sqrt{3K}$, where $|\beta|_{\max}$ is the largest coupling reached since recombination, and therefore $|\beta|_{\max}\geq\bmin$.
The coupling may be small at recombination, but it must reach $\bmin$ before the crossing.

The size of $\bmin$ can be understood in the matter era.
There dark energy is negligible, so that $\mathcal{B}\simeq0$, the CDM makes up a constant fraction $f_c\equiv\omega_c/\omega_m$ of the matter, and the mass loss is still small.
Equation~\eqref{eq:Dode} with the positive sign then becomes
\begin{equation}
\Delta'\simeq|\beta|\sqrt{3f_c\,\Delta}
\,\,\implies\,\,\Delta_{\max}\simeq\frac34\,\beta^2f_c\,(N-N_*)^2,
\end{equation}
so that the fastest mass loss grows as the square of the number of e-folds since recombination.
Requiring $\Delta_{\max}\geq\Delta_{\min}$ at every redshift gives
\begin{equation}
\bmin\simeq\max_z\frac{2}{N(z)-N_*}\sqrt{\frac{\Delta_{\min}(z)}{3\,f_c}}\,,
\label{eq:estimate}
\end{equation}
where the maximum is taken over the redshifts covered by the data.
This estimate lies about 10\% below the exact value, but it shows the two features that matter.
First, a coupling of a few per cent is enough, because the mass loss is spread over the $\simeq6.4$ e-folds between recombination and $z\simeq0.8$.
Second, $\bmin$ grows as the square root of the depth of the crossing, that is, of the peak value of $\Delta_{\min}$.

The bound is sharp: there are histories for which the coupling equals $\bmin$.
Suppose the coupling is constant, and the field rolls in one direction after recombination until it comes to rest at some redshift inside the observed range.
Up to that point the field moves at the speed limit, so its mass history is the fastest one, $\Delta_{\max}$, for its own coupling. Where the field is at rest, its kinetic energy vanishes, and the identity in Eq.~\eqref{eq:identity} gives exactly $\Delta=\Delta_{\min}$.
A smaller coupling would give less mass loss at that point, below the minimum required, so in this case the coupling is exactly $\bmin$.

The same reasoning tells us where the minimal field comes to rest.
For the coupling $\bmin$, the fastest mass history touches $\Delta_{\min}$ without crossing it, as shown in Fig.~\ref{fig:bg}(b).
At the contact point the field is at rest, so the mass stops changing for an instant and the curve is flat there.
A flat curve can touch $\Delta_{\min}$ from above only where $\Delta_{\min}$ is also flat, that is, at its peak.
The field of the minimal solution is therefore at rest where the crossing is deepest, and not at the crossing itself, where $\Delta_{\min}=0$.
After that point the field can either continue in the same direction or turn around, which corresponds to a change of sign of the energy transfer; both reproduce the same $H(z)$, with different potentials.

\subsection{Before recombination and before the data}
\label{sec:early}

The bound rests on two assumptions about the early Universe: that the observer normalises the CDM density at $z_*$, and that between recombination and the data the expansion history follows the fitted template. We check both with the full dynamics of the coupled field.

Before dark energy becomes important, only the coupling moves the field, which follows the known attractor solutions, $\phi'=-\tfrac32\beta\Omega_c$ during radiation domination~\cite{Antusch:2026ldp} and $\phi'=-2\beta\Omega_c$ during matter domination~\cite{Amendola:1999er}.
Along them the CDM mass changes by an amount of order $\beta^2$; for $\beta=0.05$ it is 0.27\% larger at matter--radiation equality, and 0.5\% larger at $z=10^6$, than at recombination.
As a consequence, an uncoupled analysis of the CMB infers a CDM density slightly different from its value at recombination.
Matching an uncoupled model to the exact early solution, by requiring the same redshift of equality, the same sound horizon or the same expansion rate at recombination, we find a difference of $+0.07$--$0.36\%$, of order $\beta^2$, which lowers $\bmin$ by about 5\%.

The second assumption matters more, because BAO and SN measure $H(z)$ only at $z\lesssim2.5$, and the field may behave differently at higher redshift. If the potential becomes important only below some redshift $z_d$, the field stays on the attractor until $z_d$. 
On the attractor the mass loss grows only linearly with the number of e-folds, while $\Delta_{\max}$ grows quadratically and overtakes it after about 2.7 e-folds ($z\simeq75$), so a potential that acts only at late times requires a larger coupling. 
In the opposite direction, if the field itself carries a small fraction of the energy before the data range, as a tracking field does, then $\mathcal{B}>0$ there: the field already has kinetic energy, rolls faster, and a smaller coupling is enough. 
To find the smallest possible $\bmin$, we add above $z=2.5$ the largest such component allowed by Planck, $\Omega_e=0.0036$ at 95\% CL~\cite{Planck:2015bue} (see also Refs.~\cite{Perivolaropoulos:2021jda,DiValentino:2021izs}), which changes the angular scale of the sound horizon by only about $4\times10^{-5}$.

We test the whole argument on forward-integrated coupled quintessence models in Sec.~\ref{sec:forward}.

%=======================================================================
\section{Results for DESI DR2}
\label{sec:results}

\subsection{Best fits and posterior distributions}
\label{sec:bestfits}
\label{sec:squeeze}

We use the public DESI DR2 Cobaya chains for flat $w_0w_a$CDM with free neutrino mass\footnote{\url{https://data.desi.lbl.gov/public/papers/y3/bao-cosmo-params/cobaya/base_mnu_w_wa/}}, which combine DESI DR2 BAO with Planck 2018 low-$\ell$ TT and EE, Planck PR4 CamSpec high-$\ell$ TT, TE and EE, Planck and ACT DR6 lensing, and one SN sample among Pantheon+, Union3 and DESY5.
We first discuss a single background for each SN sample, given by the posterior means of the chains, which we call the best fit in the following, and then the full posterior distributions.

\begin{figure}[t]
\centering
\includegraphics[width=\columnwidth]{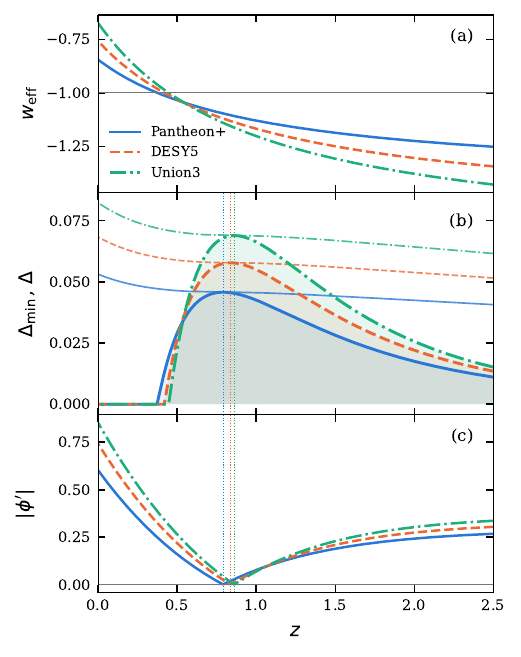}
\caption{DESI DR2 best-fit $w_0w_a$ backgrounds. (a) Apparent dark-energy equation of state. (b) Minimum mass loss $\Delta_{\min}$ (thick; shaded: forbidden region) and mass loss on the minimal-coupling trajectory (thin), which touches $\Delta_{\min}$ at its maximum. (c) Field speed on the same trajectory; it vanishes at $z_{\rm turn}\simeq0.8$, where $\Delta_{\min}$ is largest, and not at the crossing $w_{\rm eff}=-1$, where $\Delta_{\min}=0$.}
\label{fig:bg}
\end{figure}

The results for the best fits are shown in Fig.~\ref{fig:bg} and Table~\ref{tab:bf}.
The apparent crossing $w_{\rm eff}=-1$ occurs at $z\simeq0.37$--$0.45$, and its depth is $0.046$--$0.069$: by $z\simeq0.8$ the dark matter must have lost at least 4.6--6.9\% of its mass.
The resulting minimum coupling is $\bmin=0.047$, $0.052$ and $0.058$ for Pantheon+, DESY5 and Union3, about 10\% above the matter-era estimate in Eq.~\eqref{eq:estimate}.

On the minimal trajectory of Fig.~\ref{fig:bg}, the field continues in the same direction after $z\simeq0.8$, and the mass loss keeps growing, so that today the CDM density is 5--8\% below its CMB extrapolation.

\begin{table}[t]
\small
\begin{ruledtabular}
\begin{tabular}{lccc}
DESI+CMB+ & Pantheon+ & DESY5 & Union3\\
\hline
$w_0$ & $-0.845$ & $-0.758$ & $-0.674$\\
$w_a$ & $-0.570$ & $-0.820$ & $-1.057$\\
$\Omega_m$ & 0.311 & 0.319 & 0.327\\
$z_\times$ (crossing) & 0.37 & 0.42 & 0.45\\
$\max\Delta_{\min}$ & 0.046 & 0.058 & 0.069\\
at redshift & 0.79 & 0.84 & 0.86\\
$\bmin$ & 0.047 & 0.052 & 0.058\\
Eq.~\eqref{eq:estimate} & 0.042 & 0.048 & 0.052\\
$2\bmin^2$ & 0.43\% & 0.55\% & 0.66\%\\
$\Delta(0)$ & 0.053 & 0.068 & 0.082\\
\end{tabular}
\end{ruledtabular}
\caption{Results for the DESI DR2 $w_0w_a$ backgrounds, with the field at rest at recombination. The backgrounds use the posterior means of the public DESI DR2 chains. $\Delta(0)$ is the mass loss today on the minimal trajectory.}
\label{tab:bf}
\end{table}

\begin{table}[t]
\begin{ruledtabular}
\begin{tabular}{lccc}
DESI+CMB+ & Pantheon+ & DESY5 & Union3\\
\hline
baseline              & 0.046 & 0.052 & 0.057\\[2pt]
late, $z_d=2.5$ & 0.063 & 0.071 & 0.078\\
\phantom{late, }$z_d=10$  & 0.055 & 0.062 & 0.067\\
\phantom{late, }$z_d=100$ & 0.044 & 0.049 & 0.054\\
\phantom{late, }$z_d=300$ & 0.040 & 0.046 & 0.050\\[2pt]
Planck limit          & 0.034 & 0.040 & 0.045\\
\end{tabular}
\end{ruledtabular}
\caption{$\bmin$ for different histories before the data range, for the DESI DR2 best fits. Baseline: field at rest at recombination. Late: the potential is negligible above $z_d$, and the field follows the coupling-driven attractor until $z_d$ (normalisation offset included). Planck limit: largest tracking early dark energy allowed by Planck, $\Omega_e=0.0036$, above $z=2.5$. Here the background uses a fixed baryon fraction, so the baseline differs from Table~\ref{tab:bf} by less than $5\times10^{-4}$.}
\label{tab:early}
\end{table}

Table~\ref{tab:early} shows how these values depend on the history before the data range. If the potential becomes important only below $z_d=2.5$, as in thawing models, the field has less time to roll and the minimum coupling rises to $0.063$--$0.078$. 
If instead the potential acts earlier, the field has more time, and for $z_d\gtrsim100$ the minimum coupling falls 6--13\% below the baseline, because the field is already rolling on the attractor when the potential takes over. 
Finally, the largest early expansion allowed by Planck lowers the baseline by 21--26\%.

To include the uncertainties of the data, we repeat the calculation for the full posteriors. 
After removing 30\% of each chain as burn-in, we take 4000 samples and compute $\Delta_{\min}(z)$ and $\bmin$ for each of them. 
Since $|\beta|_{\max}\geq\bmin$, the 5th percentile of the distribution of $\bmin$ is a 95\% lower limit on the coupling. As Table~\ref{tab:data} shows, the coupling must be larger than $0.025$, $0.033$ and $0.038$ at 95\% CL for Pantheon+, DESY5 and Union3, and the medians agree with the best-fit values. 
The field of the minimal solution is at rest at $z\simeq0.8\pm0.1$. These limits assume that the crossing is real and due to the coupling, and they should not be read as evidence for a coupling: in the $w_0w_a$ parametrisation almost every sample with $w_a<0$ crosses $w=-1$, and fewer than 0.3\% of the samples need no coupling at all.

\begin{table}[t]
\small\setlength{\tabcolsep}{4pt}
\begin{ruledtabular}
\begin{tabular}{lccc}
DESI+CMB+ & Pantheon+ & DESY5 & Union3\\
\hline
$\bmin$                         & $0.046^{+0.012}_{-0.012}$ & $0.052^{+0.011}_{-0.012}$ & $0.057^{+0.012}_{-0.011}$\\[3pt]
95\% lower limit                & 0.025 & 0.033 & 0.038\\
\quad conservative              & 0.013 & 0.021 & 0.026\\[3pt]
$P(\bmin>0.073)$                & 1.0\% & 2.7\% & 8.1\%\\
$\max_z\Delta_{\min}$ (median)  & 0.045 & 0.057 & 0.068\\
$z_{\rm turn}$                  & $0.79^{+0.16}_{-0.11}$ & $0.84^{+0.14}_{-0.09}$ & $0.86^{+0.12}_{-0.10}$\\[3pt]
$\delta_L$ [\%]                 & $3.8^{+2.2}_{-1.8}$ & $4.8^{+2.2}_{-2.0}$ & $5.7^{+2.5}_{-2.1}$\\
\quad 95\% lower limit          & 1.0 & 1.8 & 2.5\\
\end{tabular}
\end{ruledtabular}
\caption{Distribution of $\bmin$ and related quantities from the DESI DR2 $w_0w_a$ chains: medians with 68\% intervals, and 95\% lower limits. Conservative: with the largest early expansion allowed by Planck. $z_{\rm turn}$: redshift where the minimal-coupling field is at rest. $\delta_L$: minimum lensing deficit at $z\simeq0.8$ (Sec.~\ref{sec:obs}).}
\label{tab:data}
\end{table}

\begin{figure}[t]
\centering
\includegraphics[width=\columnwidth]{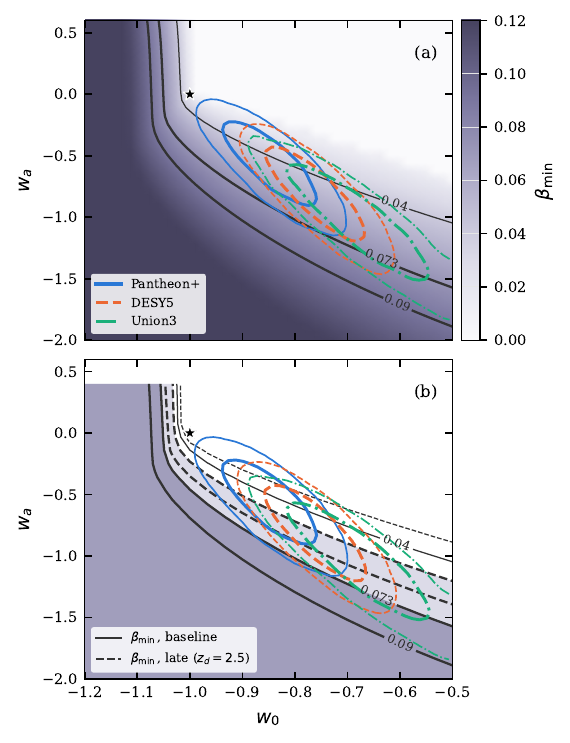}
\caption{(a) Minimum coupling $\bmin$ in the $w_0$--$w_a$ plane for the baseline history (colour and solid contours), at fixed $\Omega_m=0.315$ and $h=0.675$, with the 68\% and 95\% contours of the DESI DR2 posteriors; the star marks $\Lambda$CDM. (b) Same plane, with $\bmin$ for the baseline (solid) and late-potential (dashed, $z_d=2.5$) histories compared with the CMB upper limits on a constant coupling. Dark shading: $\bmin$ exceeds the upper limit $0.073$ for the baseline history; light shading: only for the late-potential history.}
\label{fig:map}
\end{figure}

Figure~\ref{fig:map}(a) shows where these values come from. It gives $\bmin$ across the $w_0$--$w_a$ plane, at fixed $\Omega_m$ and $h$, together with the DESI DR2 posteriors.
The minimum coupling vanishes only where $w_{\rm eff}\geq-1$ at all $z\leq2.5$, that is, for $w_0\geq-1$ and $w_0+0.71\,w_a\geq-1$, and it grows as the crossing becomes deeper, away from $\Lambda$CDM.
The DESI DR2 posteriors lie almost entirely in the region where a coupling is required, mostly between $\bmin\simeq0.03$ and $0.07$, in agreement with the 68\% intervals of Table~\ref{tab:data}.

The CMB, on the other hand, limits a constant coupling from above: $|\beta|<0.073$ at 95\% CL from fits of specific models~\cite{Babu:2026tfr}, and $|\beta|\lesssim0.09$ from the conditions at recombination of Ref.~\cite{Antusch:2026ldp} (see also Ref.~\cite{Amendola:1999er}).
The coupling is therefore squeezed between two bounds, as shown in Fig.~\ref{fig:map}(b).
In the light region the lower bound exceeds the upper one if the potential acts only at late times, and in the dark region it does so for the baseline history as well.
For a late-acting potential this happens in 33--70\% of the DESI DR2 posterior, and for the baseline history in only 1--8\%.
With the looser limit $|\beta|\lesssim0.09$, the late-potential fraction is still 8--32\%.
There, a constant coupling cannot produce the crossing with a late-acting potential: the crossing requires either a potential that acts soon after recombination or a coupling that grows towards low redshift.

\subsection{Forward model}\label{sec:forward}
We tested the whole argument on coupled quintessence models integrated forward from $z=10^5$, with a Peebles--Ratra potential, a constant coupling and the field at rest at the start, tuning the amplitude of the potential and the CDM density to $h=0.675$ and to the Planck CDM density at recombination.
Out of 240 models, with $\beta$ from $\pm0.02$ to $\pm0.08$ and different slopes and initial values of the field, 111 have $w_{\rm eff}<-1$ somewhere at $z\leq2.5$.
In all of them Eq.~\eqref{eq:identity} holds to $10^{-3}$ and neither bound is violated.
To mimic the data, we then computed $\bmin$ from the $H(z)$ of each model at $z\leq2.5$ only, with the late history of Sec.~\ref{sec:early} above.
The saturation argument of Sec.~\ref{sec:bmin} predicts that the bound is reached, $\bmin=|\beta|$, when the field comes to rest within the data range, and that otherwise $\bmin$ is smaller than the true coupling. The models confirm this. In the 28 models in which the field turns around at $z\leq2.5$, the bound recovers the true coupling to within 2.5\%, and the agreement is better the later the turnaround. In the other models, where the field rolls in one direction, the bound is on average about 12\% below the true coupling.

These models also show how deep a crossing a given coupling can produce. With $|\beta|\simeq0.03$--$0.05$, the depth of the crossing is 0.5--2.4\%, about 10--50\% of that of the DESI $w_0w_a$ best fits, and a depth comparable to DESI is reached only for $|\beta|\gtrsim0.08$. This is expected, because the Peebles--Ratra potential acts mainly at late times, so the field has little time to roll.

%=======================================================================
\section{Observable consequences}
\label{sec:obs}
The minimum coupling has two consequences beyond the background, and both have a minimum size.
The first is a fifth force.
Below the Compton wavelength of the field, two CDM particles attract each other with an effective Newton constant larger by a factor $1+2\beta^2$ ($1+\tfrac43\beta^2$ in the convention of Refs.~\cite{Amendola:1999er,Amendola:2003wa}, whose coupling is larger than ours by a factor $\sqrt{3/2}$), while baryons feel normal gravity~\cite{Amendola:1999er,Amendola:2003wa}.
The bound therefore implies a fifth force between dark matter particles of at least $2\bmin^2\simeq0.4$--$0.7\%$ of gravity at the best fits, and $0.1$--$0.3\%$ at the 95\% lower limits, or $0.03$--$0.13\%$ with the conservative limits.
This is the quantity probed by tests of the Euler equation for dark matter, whose current sensitivity is a few per cent and is expected to reach $\sim2\%$ with Stage-IV surveys~\cite{Amendola:2016saw,Bonvin:2018ckp,Castello:2024lhl,Grimm:2025pmq}.
Forecasts for relativistic multi-tracer tests of the equivalence principle in coupled dark energy find that DESI and Euclid cannot reach couplings of this size, while SKA2 could approach them~\cite{Kegel:2026ppi}.

The second consequence follows directly from the mass loss.
The gravitational potential of a CDM overdensity is proportional to its mass, so for a given clustering of CDM the lensing potential is smaller by the fraction of mass that has been lost~\cite{LaPenna:2026avs}.
Weighted by the CDM fraction of matter, the lensing per unit clustering is reduced by at least $\delta_L=f_c\Delta_{\min}$ at $z\simeq0.8$, close to the peak of the lensing kernels.
This gives $\delta_L\simeq 3.8$--$5.8\%$ at the DESI DR2 best fits and, from the distributions in Table~\ref{tab:data}, at least $1.0$--$2.5\%$ at 95\% CL.

The statement does not extend to $z=0$, because below $z\simeq0.8$ the field may turn around and part of the mass can come back.
It is, however, independent of how the growth of structure is modified, and it can be tested by comparing lensing with clustering and redshift-space distortions~\cite{Amendola:2007rr, Sapone:2010uy, Zhong:2023how,Keil:2026cat}.

The growth of structure, in contrast, is not fixed by the bound. 
It depends on $H(z)$, which the data fix, but also on the coupling and on the mass history $\Delta(z)$, which the data do not: the same $H(z)$ is produced by any $|\beta|\ge\bmin$, and the field may continue or turn around after $z\simeq0.8$. 
The fifth force and the lower friction make the growth faster, while the lower CDM density makes it slower~\cite{LaPenna:2026avs}. 
On our minimal trajectory, computed in the quasi-static limit with the same $H(z)$ and primordial amplitude as the uncoupled $w_0w_a$ model, $f\sigma_8$ at $z=0.5$ is $4.0$--$5.9\%$ lower and the lensing potential $6$--$9\%$ lower. 
Other trajectories with the same $H(z)$ can change the size and even the sign of these effects, so they are not a robust prediction (see Sec.~\ref{sec:compare}).

%=======================================================================

\section{Comparison with published fits and previous work}
\label{sec:compare}

Since the bound holds for any potential, every published fit of coupled quintessence must respect it.
Comparing with these fits is therefore both a test of the bound and a way to understand them.
To make the comparison, we note that our $\beta$ equals the coupling of Refs.~\cite{Gomez-Valent:2026ept,Li:2026xaz} up to a sign, $\lambda/\sqrt{8\pi}$ in Refs.~\cite{Antusch:2026ldp,Chakraborty:2025syu}, and $\alpha/\phi$ for $m\propto\phi^\alpha$~\cite{Pompeu:2026zad}.
We selected six best fits from four analyses~\cite{Gomez-Valent:2026ept,Li:2026xaz,Antusch:2026ldp,Pompeu:2026zad} that give all the parameters needed to reproduce their background.
They are the DES-Dovekie best fits of Ref.~\cite{Gomez-Valent:2026ept} with $\beta>0$ and $\beta<0$ and the best fit of Ref.~\cite{Li:2026xaz}, all with an exponential mass and a Peebles--Ratra potential; the fits of Ref.~\cite{Antusch:2026ldp} with an exponential mass and an inverse power-law potential, and with a quadratic mass and a linear potential; and the CMB best fit of Ref.~\cite{Pompeu:2026zad}, with $m\propto\phi^\alpha$ and a Peebles--Ratra potential.
Other analyses find similar couplings, $|\beta|=0.045$--$0.049$~\cite{Wang:2026vqw,Wang:2026gvg} (with $\Sigma m_\nu=0.06$\,eV), $0.050$~\cite{Samanta:2025oqz}, $0.05\pm0.01$ in the dark-dimension scenario~\cite{Bedroya:2025fwh}, and $\simeq0.05$ with $|\beta|<0.073$~\cite{Babu:2026tfr}, but they do not give enough information for this test.

For each of the six fits, we integrated the model forward with its values of $h$ and $\omega_c$ and applied our analysis to the resulting $H(z)$; we reproduce only the background, not the full fits.
Because BAO and SN measure $H(z)$ only at $z\lesssim2.5$, we compute $\bmin$ from the model's $H(z)$ in this range, with the field either at rest at $z_*$ (base) or on the attractor above $z=2.5$ (late).
The second history is the appropriate one for these models, whose potentials act only at late times.
We do not use the model’s $H(z)$ at all redshifts because with a constant coupling the field follows the attractor during the matter era, so the high-redshift $H(z)$ already depends on $\beta$, and the bound would return $\bmin\simeq|\beta|$ for every model, whatever happens at the crossing.

\begin{table}[t]
\footnotesize
\begin{ruledtabular}
\setlength{\tabcolsep}{2.5pt}
\begin{tabular}{cccccl}
 & depth & & \multicolumn{2}{c}{$\bmin$} & \\
$|\beta|$ & (\%) & $z_{\rm turn}$ & base & late & Ref.\\
\hline
0.056 & 2.3  & ---  & 0.034 & 0.050 & \cite{Gomez-Valent:2026ept}, $\beta>0$\\
0.029 & 0.7  & 1.8  & 0.019 & 0.028 & \cite{Gomez-Valent:2026ept}, $\beta<0$\\
0.053 & 2.2  & ---  & 0.033 & 0.049 & \cite{Li:2026xaz}\footnote{Initial field not given; $\phi_i=1$--3 changes the results by less than 2\%.}\\
0.060 & 3.1  & 1.8  & 0.040 & 0.059 & \cite{Antusch:2026ldp}, exp.\\
0.18\footnote{Value today; largest value before the peak of $\Delta_{\min}$: 0.29.} & 11.9 & 1.0 & 0.078 & 0.114 & \cite{Antusch:2026ldp}, quad.\\
0.043\footnote{$\alpha/\phi$ at the initial field; largest value before the peak of $\Delta_{\min}$: 0.047. Initial scale factor not given ($10^{-5}$ assumed).} & 1.8 & 1.6 & 0.030 & 0.044 & \cite{Pompeu:2026zad}\\
\hline
---\footnote{Local coupling 0.16--0.46.} & 5.7 & 0.8 & 0.052 & --- & \cite{LaPenna:2026avs}\\
\hline
--- & 4.6--6.9 & 0.8 & 0.047--0.058 & 0.067--0.082 & DESI DR2 $w_0w_a$\\
\end{tabular}
\end{ruledtabular}
\caption{Published coupled-quintessence best fits and the bound applied to their expansion history at $z\leq2.5$. $|\beta|$: fitted coupling; depth: $\max_z\Delta_{\min}$; $z_{\rm turn}$: turnaround redshift (--- if none); $\bmin$: field at rest at $z_*$ (base) or on the attractor above $z=2.5$ (late), without the normalisation offset, for the fits and for the DESI DR2 best fits alike. The models are described in Sec.~\ref{sec:compare}.}
\label{tab:lit}
\end{table}

\begin{figure}[t]
\centering
\includegraphics[width=\columnwidth]{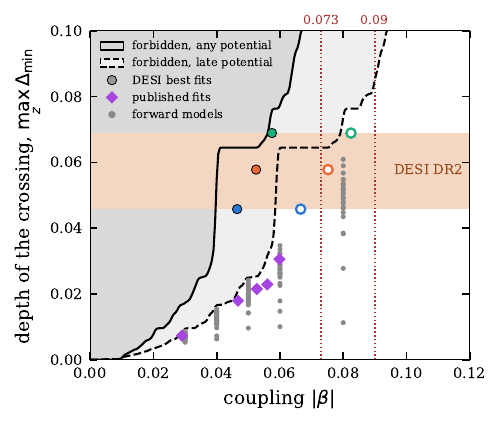}
\caption{Deepest apparent crossing that a coupling $|\beta|$ can produce, from the $w_0w_a$ histories of Fig.~\ref{fig:map}. Dark grey: forbidden for any potential; light grey: forbidden for potentials acting only at $z<2.5$. Band and circles: DESI DR2 best fits (filled: baseline; open: late potential). Diamonds: published fits; small points: forward models; dotted lines: CMB upper limits.}
\label{fig:depth}
\end{figure}

The results, collected in Table~\ref{tab:lit} and Fig.~\ref{fig:depth}, tell a consistent story.
All the fits respect the bound.
For the constant couplings, $\bmin$ is 61--66\% of the fitted $|\beta|$ with the base history and 90--98\% with the late one.
In the three fits in which the field turns around in the data range it reaches 95--98\%, so these fits saturate the bound, as Sec.~\ref{sec:bmin} predicts.
For the quadratic coupling of Ref.~\cite{Antusch:2026ldp}, which grows at late times, the ratio drops to 0.39, simply because the bound applies to the largest value of the coupling.
The most telling number, however, is the depth.
The constant-coupling fits produce crossings of only 0.7--3.1\%, well below the 4.6--6.9\% of the DESI $w_0w_a$ best fits, and their couplings are close to the minimum ones for these shallower crossings.
This is consistent with Refs.~\cite{Gomez-Valent:2026ept,Babu:2026tfr,Li:2026xaz}, which find effective equations of state that evolve less than $w_0w_a$, or recover only part of its improvement over $\Lambda$CDM.
Only the growing coupling of Ref.~\cite{Antusch:2026ldp}, small at recombination and $0.18$ today, produces a deep crossing (11.9\% at $z\simeq1$), as the squeeze of Sec.~\ref{sec:squeeze} suggests.

Several elements of our analysis have appeared before in other forms.
Ref.~\cite{LaPenna:2026avs} showed, for one $w_0w_a$ background, that the dark matter mass must avoid a forbidden region; Eq.~\eqref{eq:Dmin} gives this region for any background, and for their reconstruction we find $\bmin=0.052$, while their local coupling reaches 0.16--0.46.
Refs.~\cite{Pompeu:2026zad,Wang:2026wrk} noted that the crossing involves a turnaround of the field, or a change of sign of the energy transfer, which is one of the saturation cases discussed above.
Some works find $|\beta|\simeq0.03$--$0.05$ even without a crossing~\cite{Gomez-Valent:2026ept,Li:2026xaz,Wang:2026vqw}, which shows that the data can prefer a coupling of this size for other reasons, for example through the dark matter density.
Our bound does not explain these values; it gives the coupling that a given crossing requires.

The minimal solution also has a simple interpretation in terms of the field excursion.
On the minimal trajectory the field moves by $|\Delta\phi|\simeq\max_z\Delta_{\min}/\bmin\simeq1.0$--$1.2$ in Planck units between recombination and the turnaround, and larger couplings give smaller excursions.
The apparent crossing is therefore compatible with the swampland distance conjecture~\cite{Ooguri:2006in}, which disfavours excursions much larger than the Planck mass, for $|\beta|\gtrsim\bmin$.
A coupling of this size also appears in the dark-dimension scenario~\cite{Bedroya:2025fwh}.

The same reasoning applies to coupled dark-energy fluids with $w\geq-1$, for which Eq.~\eqref{eq:Dmin} still holds.
For $Q=\beta H\rho_c$ the mass loss after $N$ e-folds is $1-e^{-\beta N}$, and the bound requires $\beta\geq0.007$--$0.011$.
The CMB instead gives $\beta=0.0030\pm 0.0014$~\cite{Li:2026jgg}, 3--6$\sigma$ lower, in agreement with the weak evolution of $w_{\rm eff}$ found in that work.
The fluid models that do reproduce the crossing use a fluid with $w\simeq-1.5$~\cite{Li:2026jgg,Li:2026xaz}, outside our assumptions.

Finally, the observable consequences of Sec.~\ref{sec:obs} can be compared with existing analyses.
Ref.~\cite{Clini:2026kdj} finds a $2.5$--$3\sigma$ hint of a dark-sector force of $\simeq0.5\%$ of gravity in CMB and DESI clustering data, of the same size as our minimum fifth force, although in an analysis without evolving dark energy.
For the growth of structure, Refs.~\cite{LaPenna:2026avs,Khoury:2025txd} find a suppression, as on our minimal trajectory, whereas full fits that vary all the parameters find an enhancement, with $\sigma_8$ about $2\sigma$ higher~\cite{Pompeu:2026zad,Ganguly:2026mgu}, as expected from the trajectory dependence discussed in Sec.~\ref{sec:obs}.

%=======================================================================
\section{Conclusions}
\label{sec:conclusions}

If the crossing of $w=-1$ preferred by DESI is due to a quintessence field coupled to dark matter, its depth fixes a minimum coupling, for any potential.
The result rests on two facts.
The field can appear phantom only if the dark matter loses mass, so the expansion history fixes a minimum mass loss since recombination.
The dark matter can lose mass only as fast as the field moves, and the speed of the field is itself fixed by the expansion history, so a coupling that is too small cannot remove enough mass in time.
The bound is reached when the field comes to rest within the data range, and for the DESI backgrounds this happens at $z\simeq0.8$, where the crossing is deepest, and not at the crossing itself.

For the DESI DR2 best fits the dark matter must lose at least 5--7\% of its mass after recombination, and the coupling must be at least $0.047$--$0.058$.
Including the uncertainties of the data, the public chains give $|\beta|>0.025$--$0.038$ at 95\% CL, and $|\beta|>0.013$--$0.026$ even with the largest early expansion allowed by Planck.
The requirement depends on the history before the data: if the field is frozen until $z\simeq2.5$, as in thawing models, it rises to $0.06$--$0.08$, at the level of the CMB upper limits on a constant coupling.
For such a late-acting potential, a constant coupling compatible with the CMB cannot produce the crossing in 33--70\% of the DESI DR2 posterior.
A potential that acts soon after recombination, or a coupling that grows at late times, is then the natural way to reproduce the full crossing.

We tested the argument in two ways.
In coupled quintessence models integrated forward, the bound is never violated, and it recovers the true coupling whenever the field turns around within the data range.
Published fits with a constant coupling also respect the bound, but their expansion histories require a mass loss of only $0.7$--$3.1$\%, against $4.6$--$6.9$\% for the DESI $w_0w_a$ best fits; their crossings are shallower, which explains why they recover only part of the DESI signal.
The same mass-loss argument applies to coupled dark-energy fluids with $w\geq-1$, for which the coupling required by the crossing is several standard deviations above the value allowed by the CMB.

The bound also gives two observable effects a minimum size.
The dark matter particles must feel a fifth force of at least $0.4$--$0.7\%$ of gravity at the best fits, and $0.1$--$0.3\%$ at 95\% CL, below the $\sim2\%$ sensitivity expected for Stage-IV tests of the dark-matter Euler equation, although SKA2-like surveys could approach it~\cite{Kegel:2026ppi}.
The lensing per unit clustering must be lower by $3.8$--$5.8\%$ at $z\simeq0.8$, and by at least $1.0$--$2.5\%$ at 95\% CL, an effect within reach of the comparison of lensing with clustering in DESI~\cite{DESI:2016fyo}, Euclid~\cite{ Euclid:2024yrr,Euclid:2021osj}, and Rubin Observatory~\cite{LSSTDarkEnergyScience:2018jkl}.
If such a deficit is not found, canonical coupled quintessence will be excluded as the origin of the crossing, whatever the potential.
Our bound applies to canonical scalar fields coupled to dark matter through its mass; momentum exchange, non-canonical kinetic terms and nonminimal couplings to gravity~\cite{Wolf:2025jed,Ye:2024ywg,Tsujikawa:2025wca} would require a separate study, as would a non-parametric reconstruction of $\mathcal{B}(z)$ from future BAO and SN data.

\begin{acknowledgments}
We thank Savvas Nesseris and Spyros Sypsas for useful discussions and comments on the draft.
DS acknowledges support from the Fondecyt Regular project 1251339 (ANID). EA acknowledges support by the EPEC Scholarship program from the Universidad de Chile.
\end{acknowledgments}

%=======================================================================
\bibliography{references}

\end{document}